\documentclass[aps,prd,reprint,nofootinbib,bibnotes,groupedaddress]{revtex4-2}
\usepackage{amsmath,bbm,hyperref,mathrsfs}
\begin{document}

\title{Isospectral Local Hamiltonians for Perturbative $\mathcal{PT}$-symmetric Hamiltonians}

\author{Yi-Da Li}
\email{liyd20@mails.tsinghua.edu.cn}
\affiliation{Department of Physics, Tsinghua University, Beijing 100084, P. R. China}

\author{Qing Wang}
\email[Corresponding author:~]{wangq@mail.tsinghua.edu.cn}
\affiliation{Department of Physics, Tsinghua University, Beijing 100084, P. R. China \\
Center for High Energy Physics, Tsinghua University, Beijing 100084, P. R. China}

\date{\today}

\begin{abstract}
A new method to work out the Hermitian correspondence of a $\mathcal{PT}$-symmetric quantum mechanical Hamiltonian is proposed. In contrast to the conventional method, the new method ends with a local Hamiltonian of the form $\frac12p^2+\frac{1}{2}m^2x^2+v(x)$ without any higher-derivative terms. This method is demonstrated in the perturbative regime. Possible extensions to multi-variable quantum mechanics and quantum field theories are discussed.
\end{abstract}

\maketitle

\section{Introduction\label{sec-intro}}

The discovery of real-spectra $\mathcal{PT}$-symmetric Hamiltonians\cite{bender1998} has inspired a lot of researches beyond conventional Hermitian quantum theories\cite{bender-review,bender-book}. Originally, in \cite{bender1998} it was found that Hamiltonians of the form $H=p^2+m^2-(ix)^{N} (N\ge2)$ have real spectra. Later, the general framework to describe a $\mathcal{PT}$-smmetric quantum theory was established\cite{bender2002,ali2002-1,ali2002-2,ali2002-3}. A nontrivial metric operator $\eta=e^{-Q}$ satisfying $\eta H\eta^{-1}=H^\dagger$ is necessary\cite{ali2002-1} for the unitary evolution generated by a non-Hermitian $\mathcal{PT}$-symmetric Hamiltonian $H$, which differs from Hermitian quantum mechanics. With the help of this metric operator, a real-spectra $\mathcal{PT}$-symmetric Hamiltonian $H$ can be recast to an isospectral Hermitian Hamiltonian $h=e^{-Q/2}He^{Q/2}$ equipped with the ordinary Dirac inner product. A remarkable example is the isospectral Hermitian Hamiltonian for $H=p^2-gx^4$, as described in \cite{jones2006,bender2006}. The stability for $-x^4$ potential is essential to guarantee the stability of Higgs vacuum\cite{bender-book}. Moreover, a generic method\cite{bender2004} has been developed to calculate the metric operator for a perturbative $\mathcal{PT}$-symmetric Hamiltonian of the form $H=H_0+\epsilon H_1$ where $H_0$ is Hermitian and $H_1$ is anti-Hermitian. In this case, $Q$ has the form $Q=\epsilon Q_1+\epsilon Q_3+\cdots$ and each term can be determined perturbatively as follows\cite{bender-book}
\begin{equation}
[H_0,Q_1]=-2H_1,\ [H_0,Q_3]=-\frac16[[H_1,Q_1],Q_1],\ \cdots
\end{equation}
The isospectral Hermitian Hamiltonian $h$ acquired from this procedure is in general nonlocal in the sense of containing terms in arbitrarily high order of momentum $p$, which render the physical meaning of $h$ rather obscure. However, there are vast degrees of freedom in generating $h$ as demonstrated in \cite{bender2009}. In this paper we give an explicit method to calculate the local version of $h$ for perturbative $\mathcal{PT}$-symmetric Hamiltonians whose free parts are non-degenerate. In contrast to the nonlocal $h$ from above conventional method, we believe a local form has apparent physical meanings and will bring inspirations to the research of $\mathcal{PT}$-symmetric theories.

Here we summarize main procedures of our new methods and the structure of this paper. In Sec. \ref{sec-diag}, we start from a single-variable Hamiltonian $H_V=\frac12p^2+\frac12m^2x^2+V(x,p)$, where $V(x,p)=\sum_{n=1}^\infty g^nV_n(x,p)$ is the sum of various polynomial functions $V_n(x,p)$ of $x$ and $p$ with coupling constant $g$. We assume $H_V$ respects unbroken $\mathcal{PT}$ symmetry. Then we show a similarity transformation of $H_V$ leads to a manifestly diagonal Hermitian Hamiltonian $H_N=m(N+\frac12)+F(N)$, where $F(N)=\sum_{n=1}^\infty g^nf_n(N)$ is the sum of various polynomial functions $f_n(N)$ of $N$ and $N=a^\dagger a$ in which $a=\sqrt{\frac{m}{2}}x+i\sqrt{\frac{1}{2m}}p$ is the standard annihilation operator\footnote{$\hbar=1$ is assumed.}. In Sec. \ref{sec-local}, we transform $H_N$ to $h_v=\frac12p^2+\frac12m^2x^2+v(x)$, where $v(x)=\sum_{n=1}^\infty g^nv_n(x)$ is the sum of various polynomial functions $v_n(x)$ of $x$ only. The tranformation from $H_V$ to $H_N$ is a typical diagonalization procedure. And the key point of the transformation from $H_N$ to $h_v$ is the existence of a one-to-one correspondence between $n$-th order polynomials of $N$ and $x^2$. In Sec. \ref{sec-ix3}, we calculate $h_v$ in the $ix^3$ model as an example. When generalizing to multi-variable Hamiltonians, the one-to-one map exists only in the case where the free part of $H_V$ is non-degenerate, and this is discussed in Sec. \ref{sec-gen} together with generalization to quantum field theories. We conclude in Sec. \ref{sec-sum}.

\section{Diagonalization of a Hamiltonian with the $D$-operation\label{sec-diag}}

Consider a single-variable Hamitonian with one\footnote{Generalization to multi-coupling Hamiltonians is straightforward.} real coupling constant $g$
\begin{equation}
H_V=\frac12p^2+\frac12m^2x^2+\sum_{n=1}^\infty g^nV_n(x,p).
\end{equation}
As stated in Sec. \ref{sec-intro}, $V_n(x,p)$ is a polynomial function of $x$ and $p$ respecting $\mathcal{PT}$ symmetry. Creation and annihilation operators can be defined as usual
\begin{equation}
a^\dagger=\sqrt{\frac{m}{2}}x-i\sqrt{\frac{1}{2m}}p,\ a=\sqrt{\frac{m}{2}}x+i\sqrt{\frac{1}{2m}}p.
\end{equation}
In the Fock space defined by $a^\dagger$ and $a$, diagonal operators are in the form $\sum_{n=0}^\infty c_n N^n$ because of the commutation relation $[a,a^\dagger]=1$. We define a linear operation $D()$ on any operator $\mathcal{O}$ to take out the diagonal part of $\mathcal{O}$ such that $D(\mathcal{O})=\sum_{n=0}^\infty c^{\mathcal{O}}_n N^n$. For example,
\begin{equation}\label{eq-d}\begin{aligned}
&D(1)=1,\ D(x)=D(p)=0,\\
&D(x^2)=\frac{1}{m^2}D(p^2)=\frac{1}{2m}(2N+1),\cdots
\end{aligned}\end{equation}
A diagonal operator $\mathcal{O}$ satisfies $\mathcal{O}=D(\mathcal{O})$. If we want to diagonalize $H_V$ with a similarity transformation $e^{-R}$, it is enough to satisfy the condition
\begin{equation}\label{eq-trans1}
e^{-R}H_Ve^{R}=D(e^{-R}H_Ve^{R}).
\end{equation}

\begin{widetext}
Assume $H_V$ can be evaluated in the perturbative regime, then $R$ can be written as  a perturbation series $R=\sum_{n=1}^\infty g^nR_n$. Taking out $n$-th order terms on both sides of (\ref{eq-trans1}), we have
\begin{equation}\label{eq-rn}\begin{aligned}
[H_0,R_n]=&D([H_0,R_n])+D(V_n)-V_n\\
&+D\left(\sum_{j=2}^n\sum_{\stackrel{\{k_1,\cdots,k_j\}}{k_1+\cdots+k_j=n}}\frac{[[H_0,R_{k_1}],\cdots,R_{k_j}]}{j!}+\sum_{\ell=1}^{n-1}\sum_{j=1}^{n-\ell}\sum_{\stackrel{\{k_1,\cdots,k_j\}}{k_1+\cdots+k_j=n-\ell}}\frac{[[V_\ell,R_{k_1}],\cdots,R_{k_j}]}{j!}\right)\\
&-\left(\sum_{j=2}^n\sum_{\stackrel{\{k_1,\cdots,k_j\}}{k_1+\cdots+k_j=n}}\frac{[[H_0,R_{k_1}],\cdots,R_{k_j}]}{j!}+\sum_{\ell=1}^{n-1}\sum_{j=1}^{n-\ell}\sum_{\stackrel{\{k_1,\cdots,k_j\}}{k_1+\cdots+k_j=n-\ell}}\frac{[[V_\ell,R_{k_1}],\cdots,R_{k_j}]}{j!}\right),
\end{aligned}\end{equation}
where $H_0\equiv\frac12p^2+\frac12m^2x^2=m\left(N+\frac12\right)$.

Because $H_0$ is diagonal, $[H_0,R_n]$ has vanishing diagonal components and $D([H_0,R_n])=0$. $[H_0,R_n]$ is thus determined completely by lower-order $R_k$s. It is obvious that $D(\mathcal{O})-\mathcal{O}$ has vanishinig diagonal components such that it is in the form $\sum_{k,\ell(k\neq\ell)}{c'}^{\mathcal{O}}_{k\ell}a^{\dagger k}a^\ell$. We also have the relation $[H_0,a^{\dagger k}a^\ell/(m(k-\ell))+\alpha_{k\ell}(N)]=a^{\dagger k}a^\ell$ where $\alpha_{k\ell}(N)$ is an arbitrary function of $N$, so that all $R_n$ can be solved iteratively from (\ref{eq-rn}). $H_N\equiv e^{-R}H_Ve^{R}$ is thus in the form $H_N=m(N+\frac12)+\sum_{n=1}^\infty g^nf_n(N)$ as stated in Sec. \ref{sec-intro}, where $f_n(N)$ is given by
\begin{equation}\label{eq-fn}
f_n(N)=D\left(V_n+\sum_{j=2}^n\sum_{\stackrel{\{k_1,\cdots,k_j\}}{k_1+\cdots+k_j=n}}\frac{[[H_0,R_{k_1}],\cdots,R_{k_j}]}{j!}+\sum_{\ell=1}^{n-1}\sum_{j=1}^{n-\ell}\sum_{\stackrel{\{k_1,\cdots,k_j\}}{k_1+\cdots+k_j=n-\ell}}\frac{[[V_\ell,R_{k_1}],\cdots,R_{k_j}]}{j!}\right).
\end{equation}

\section{The Local Potential from a Diagonal Hamiltonian\label{sec-local}}

The diagonalization of $H_V$ makes use of the $D$-operation, and one may think that $H_V$ can be recovered from the diagonal $H_N$ by some $D^{-1}$-operation. However, the $D$-operation is not bijective as shown by (\ref{eq-d}) such that $D^{-1}$ does not exist. The non-existence of $D^{-1}$ indicates that there are many different Hamiltonians which is similar to the same diagonal $H_N$. As we are going to show, there exists a local Hermitian $h_v$ similar to $H_N$ serving as the Hermitian correspondences of $H_V$.

To invert the diagonalization procedure, we make use of the fact that $D(x^{2n})$ is a polynomial function of $N$ written as
\begin{equation}\label{eq-x}
D(x^{2n})=\sum_{k=0}^nX_{nk}N^k,
\end{equation}
where $X_{nn}\neq0$. Then we can define a linear operation $L()$ on any operator $\mathcal{O}$ as follows
\begin{equation}\label{eq-l}
L(\mathcal{O})=L(D(\mathcal{O})),\ L(1)=1, L(N^n)=\frac{1}{X_{nn}}\left(x^{2n}-\sum_{k=0}^{n-1}X_{nk}L(N^k)\right)\ (n\ge1),
\end{equation}
and $L(N^n)$ can be solved iteratively resulting a $2n$-th order polynomial function of $x$.

The requirement that $h_v\equiv e^{-K}H_Ne^{K}$ is local, is simply
\begin{equation}\label{eq-trans2}
e^{-K}H_Ne^{K}-H_0=L\left(e^{-K}H_Ne^{K}-H_0\right).
\end{equation}
Assume $K$ has a perturbative expansion $K=\sum_{n=1}^\infty g^nK_n$. Taking out $n$-th order terms on both sides of (\ref{eq-trans2}), we have
\begin{equation}\label{eq-kn}\begin{aligned}
[H_0,K_n]=&L([H_0,K_n])+L(f_n(N))-f_n(N)\\
&+L\left(\sum_{j=2}^n\sum_{\stackrel{\{k_1,\cdots,k_j\}}{k_1+\cdots+k_j=n}}\frac{[[H_0,K_{k_1}],\cdots,K_{k_j}]}{j!}+\sum_{\ell=1}^{n-1}\sum_{j=1}^{n-\ell}\sum_{\stackrel{\{k_1,\cdots,k_j\}}{k_1+\cdots+k_j=n-\ell}}\frac{[[f_\ell(N),K_{k_1}],\cdots,K_{k_j}]}{j!}\right)\\
-&\left(\sum_{j=2}^n\sum_{\stackrel{\{k_1,\cdots,k_j\}}{k_1+\cdots+k_j=n}}\frac{[[H_0,K_{k_1}],\cdots,K_{k_j}]}{j!}+\sum_{\ell=1}^{n-1}\sum_{j=1}^{n-\ell}\sum_{\stackrel{\{k_1,\cdots,k_j\}}{k_1+\cdots+k_j=n-\ell}}\frac{[[f_\ell(N),K_{k_1}],\cdots,K_{k_j}]}{j!}\right).
\end{aligned}\end{equation}
Because $[H_0,K_n]$ has vanishing diagonal components, we have $L([H_0,K_n])=0$ by using $D([H_0,K_n])=0$ and (\ref{eq-l}). $[H_0,K_n]$ is thus determined completely by lower-order $K_k$s. From (\ref{eq-x}) and $(\ref{eq-l})$ it is obvious that $D(L(\mathcal{O}))=D(\mathcal{O})$ which is to say $L(\mathcal{O})-\mathcal{O}$ has vanishing diagonal components, for any operator $\mathcal{O}$. $K_n$ can thus be solved iteratively by the same reason of $R_n$'s as in Sec. \ref{sec-diag}. From (\ref{eq-l}), $h_v$ is finally written in the form $h_v=\frac12p^2+\frac12m^2x^2+\sum_{n=1}^\infty g^nv_n(x)$ where $v_n(x)$ is given by
\begin{equation}\label{eq-vn}
v_n(x)=L\left(f_n(N)+\sum_{j=2}^n\sum_{\stackrel{\{k_1,\cdots,k_j\}}{k_1+\cdots+k_j=n}}\frac{[[H_0,K_{k_1}],\cdots,K_{k_j}]}{j!}+\sum_{\ell=1}^{n-1}\sum_{j=1}^{n-\ell}\sum_{\stackrel{\{k_1,\cdots,k_j\}}{k_1+\cdots+k_j=n-\ell}}\frac{[[f_\ell(N),K_{k_1}],\cdots,K_{k_j}]}{j!}\right).
\end{equation}

\section{$ix^3$ as an Example\label{sec-ix3}}

The $ix^3$ model is a popular toy model for studying $\mathcal{PT}$-symmetric theories\cite{bender2004,ali2005,bender-review,bender-book}. However, a local form of the isospectral Hermitian Hamiltonian has not been given yet. Here we calculate the $h_v$ for $H_V=\frac12p^2+\frac12m^2x^2+igx^3$ up to $\mathcal{O}(g^3)$ and show that $h_v$ is indeed local. Higher-order calculation is systematic as shown by (\ref{eq-rn})(\ref{eq-fn})(\ref{eq-l})(\ref{eq-kn})(\ref{eq-vn}) but rather tedious. Higher-order terms can be calculated whenever needed and will not be presented in this paper. 

Various quantities entailed in the calculation of $h_v$ for $H_V=\frac12p^2+\frac12m^2x^2+igx^3$ is as follows, up to $\mathcal{O}(g^3)$, and we take all homogeneous terms when solving for $R_n$ from (\ref{eq-rn}) to be zero.
\begin{equation}\begin{aligned}
R_1=&\frac{-i}{m(2m)^{3/2}}\left(\frac{a^{\dagger3}}{3}+3a^\dagger+3a^{\dagger2}a-3a^\dagger a^2-3a-\frac{a^3}{3}\right),\\
R_2=&\frac{1}{m(2m^4)}\left(\frac32a^{\dagger4}-12a^{\dagger2}-6a^{\dagger3}a+6a^{\dagger}a^3+12a^2-\frac32a^4\right),\\
f_1(N)=&0,\ f_2(N)=\frac{1}{8m^4}(30N^2+30N+11),\\
L(N)=&mx^2-\frac12,\ L(N^2)=\frac23m^2x^4-mx^2,\\
v_1(x)=&0,\ v_2(x)=\frac{5}{2m^2}x^4-\frac{1}{2m^4}.
\end{aligned}\end{equation}
The expression for $h_v$ is thus
\begin{equation}
h_v=\frac12p^2+\frac12m^2x^2+\frac{5g^2}{2m^2}x^4-\frac{g^2}{2m^4}+\mathcal{O}(g^3).
\end{equation}
A typical result of $h$ using the conventional method proposed in \cite{bender2004} is\cite{ali2005}
\begin{equation}
h=\frac12p^2+\frac12m^2x^2+\frac{3g^2}{2m^4}\left(\left\{x^2,p^2\right\}+m^2x^2+\frac23\right)+\mathcal{O}(g^3),
\end{equation}
where the appearance of $\left\{x^2,p^2\right\}=x^2p^2+p^2x^2$ makes the physical interpretation of $h$ rather complicated.

\section{Generalization to Multi-Variable Quantum Mechanics and Quantum Field Theories\label{sec-gen}}

Consider a multi-variable Hamiltonian with coupling constant $g$
\begin{equation}
H_V=\sum_{i}\left(\frac12p^2_i+\frac12m^2_ix_i^2\right)+\sum_{n=1}g^nV_n(\{x_j\},\{p_k\}).
\end{equation}
If there is no degeneracy in the free part $H_0=\sum_{i}\left(\frac12p^2_i+\frac12m^2_ix_i^2\right)$, which is to say that all linear combinations of integral multiple of $m_i$ in the form $\sum_{i}n_im_i\ (n_i\in\mathbbm{Z},\ \exists n_j\neq0)$ is nonzero, there is no obstacle in calculating $h_v$ from $H_V$. First, the $D$-operation is generalized trivially resulting functions of $N_i=a^{\dagger}_ia_i$, and $D(\mathcal{O})-\mathcal{O}$ is a linear combination of $\prod_{i,j}a^{\dagger n_i}_ia^{\ell_j}_j\ \left(\sum_{i,j}\left(n_im_i-\ell_jm_j\right)\neq0\right)$ for any operator $\mathcal{O}$. Second, $R_n$ is guaranteed to have solutions by the explicit commutation relation $[H_0,\left(\prod_{i,j}a^{\dagger n_i}_ia^{\ell_j}_j\right)/\left(\sum_{i,j}\left(n_im_i-\ell_jm_j\right)\right)+\alpha_{\{n_i,\ell_j\}}(\{N_k\})]=\prod_{i,j}a^{\dagger n_i}_ia^{\ell_j}_j$. Next, the $L$-operation is also generalized trivially resulting functions of $x_i$, and $K_n$ is soluble as the same as $R_n$. Finally, we get a local $h_v=\sum_{i}\left(\frac12p^2_i+\frac12m^2_ix_i^2\right)+\sum_{n=1}g^nv_n(x_i)$ as the isospectral Hermitian Hamiltonian of $H_V$.

However, if degeneracy does occur in $H_0$, $D(\mathcal{O})-\mathcal{O}$ has terms in the form of $\prod_{i,j}a^{\dagger n_i}_ia^{\ell_j}_j$ where $\sum_{i,j}\left(n_im_i-\ell_jm_j\right)=0$. Consequently, $R_n$ has no solution and the whole procedure breaks down.

A quantum field theory is multi-variable, of course. However, Lorentz symmetry requires that all relativistic quantum field theories have the same spectra as free theories. Therefore, any perturbatively well-defined relativistic quantum field theory is equivalent to its corresponding free theory up to a similarity transformation which is constructed explicitly in textbooks such as \cite{weinberg}. A $\mathcal{PT}$-symmetric relativistic quantum field theory is thus isospectral to any local Hermitian theories having the same mass, and restricting conditions other than equivalent spectra are needed to isolate a meaningful one for a $\mathcal{PT}$-symmetric relativistic quantum field theory.

\end{widetext}

\section{Summary and Outlook\label{sec-sum}}

In this paper we propose a new method to calcluate isospectral Hermitian Hamiltonians of $\mathcal{PT}$-symmetric Hamiltonians and local expressions are acquired for those whose free parts are non-degenerate. In summary, we diagonalize a quantum mechanical Hamiltonian and transform the diagonalized one into a Hermitian Hamiltonian with a local potential making use of a correspondence between $n$-th order polynomials of $N$ and $x^2$. However, this correspondence, which is denoted as $L$-operation, is not unique. There are many polynomials that lead to the same result as $x^{2n}$ under the $D$-operation because $D(x^{2k+1})=0$ is satisfied for any rational number $k$. Therefore, various definitions of $L(N^n)$ can differ by arbitrary functions of $x^{2k+1}$ thus resulting different $h_v$s which differ from each other by arbitrary functions of $x^{2k+1}$, too. This nonuniqueness reflects spectral equivalence of different potentials and disappears once we specify the parity property of $h_v$.

Our method is incapable of dealing with theories degenerate in their free parts as discussed in Sec. \ref{sec-gen}. However, the conventional method is also invalid in this case. For example, consider a $\mathcal{PT}$-symmetric Hamiltonian $H=\frac12p^2_1+\frac12p^2_2+\frac12m^2x_1^2+\frac12(2m)^2x_2^2+igx_1^2x_2$, and the first-order equation needed to calculate the metric operator $\exp\left(\sum_{n=1}^\infty g^{2n+1}Q_{2n+1}\right)$ is $[H_0,Q_1]=-2ix_1^2x_2$, which has no solution because $\langle 2,0|[H_0,Q_1]|0,1\rangle=0$ is not consistent with $-2i\langle 2,0|x_1^2x_2|0,1\rangle=-2i/(2m)^{3/2}$ where $|2,0\rangle$ and $|0,1\rangle$ are bases in the Fock space of $H_0=\frac12p^2_1+\frac12p^2_2+\frac12m^2x_1^2+\frac12(2m)^2x_2^2$. We hope more powerful methods can be developed to handle degeneracy problems.

Although degeneracy also occurs in quantum field theories, Lorentz symmetry makes all quantum field theories with the same physical mass equivalent to each other. While conventional method picks up a Hermitian $h$ for a $\mathcal{PT}$-symmetric Hamiltonian $H$ by its explicit calculation procedure, we point out that there is no special choice of $h$ if we consider only the spectrum of a $\mathcal{PT}$-symmetric Hamiltonian $H$ and further constraints must be added to select a meaningful Hermitian $h$. We hope to extract more physical information from $\mathcal{PT}$-symmetric quantum field theories thus being able to construct a special Hermitian Hamiltonian $h$ for a $\mathcal{PT}$-symmetric Hamiltonian $H$, which carries the same physical information as $H$.

\begin{acknowledgments}

\end{acknowledgments}

\bibliography{ref.bib}

\end{document}